\documentclass[12pt]{iopart}
\usepackage{graphicx}
\usepackage{rotating}

\begin{document}
\title{Generating Converging Eigenenergy Bounds for Multidimensional Systems: A New Moment Representation, Algebraic, Quantization Formalism}
\author{Carlos R. Handy}
\address{Department of Physics, Texas Southern University, Houston, Texas 77004}
\ead{carlos.handy@tsu.edu}

\begin{abstract}
For low dimension systems admitting a moment equation representation (MER), the development of an effective eigenenergy bounding theory applicable to all discrete states had remained elusive, until now. Whereas Handy et al (1988 Phys. Rev. Lett. 60 253) demonstrated the effectiveness of the {\it Moment Problem} based, Eigenvalue Moment Method (EMM), for  generating arbitrarily tight bounds to the multidimensional, positive, bosonic ground state, its extension to arbitrary excited states seemed intractable. We have discovered a new, moment representation based, quantization formalism that achieves this. Unlike EMM, no convex optimization methods are required. The entire formulation is algebraic. As a result of our preliminary investigation, we are able to match, or surpass, the excellent, but intricate, analysis of Kravchenko et al (1996 Phys. Rev. A 54 287) with respect to the quadratic Zeeman effect, for a broad range of magnetic field strengths. Unlike their analysis, the proposed method is simple, involves no truncations, and the projection of the quantum operator is exact, within each moment subspace. Our new approach, the Orthogonal Polynomial Projection Quantization-Bounding Method (OPPQ-BM), exploits the implicit bounding capabilities of a previous method developed by Handy and Vrinceanu  (2013 J. of Phys. A: Math. Theor. 46 135202). What emerges is a completely new type of analysis (i.e. constrained quadratic form minimization) that validates the importance of moment equation representations for quantizing physical systems. Whereas the underlying principles of EMM guarantee it to be more efficient than OPPQ-BM (i.e. fewer moments are required for the same level of accuracy), the ability to implement algebraic computations, as opposed to pursuing nonlinear convex optimization methods (which can be relaxed through linear programming alternatives) recommends OPPQ-BM. We give an overview of the new method with applications.
\end{abstract}
\submitto{\JPA}
\newpage
\pagestyle{plain}

\section{Introduction}

\subsection{Overview: The Importance of Bounding Methods}

It is well recognized that the development of effective bounding methods for generating tight (converging) lower and upper bounds to the discrete state energies is an important problem. This is because many systems,  particularly those exhibiting strong coupling interactions, involve significant multiscale dynamics mandating the use of singular perturbation methods [1]. These specialized methods, including the adapation of conventional methods (i.e. large order nonorthogonal basis expansions, large order perturbation resummation analysis, asymptotic analysis, etc.), may yield varying and/or inaccurate results, motivating the need for tight bounds by which to gauge the reliability of competing estimation methods. Compounding these challenges,  existing bounding methods have their respective limitations. For instance, Rayleigh-Ritz [2] can be used to generate upper bounds to the discrete state energies; however, for singular perturbation-strong coupling systems (SPSCS), unless the basis is specifically adapted to the underlying physics, there is no guarantee that the manifestly converging upper bounds are accurately approximating the physical energies. Other methods for generating lower bounds, such as those based on Temple's lower bound method [3], are too slowly convergent. An alternate approach using Barta's lower bound theorem [4], can be ineffective since once a lower bound has been generated, there is no procedure for improving it.

Only one bounding philosophy persists that has non of the limitations of the above approaches and is partcularly well suited for addressing SPSCS problems. These correspond to bounding methods that are based on the existence of a moment equation representation (MER) for the given, low dimension,  Schrodinger operator.  Quantum systems admitting a MER representation correspond to a large and important class of physical systems; therefore, many problems can be addressed. 

There are two MER based bounding methods. The first is the Eigenvalue Moment Method (EMM) discovered by Handy et al in the mid 1980s [5-7]. Its distinguishing features are summarized below. However, in its present form it is only applicable to the multidimensional bosonic ground state, despite the fact that it can be applied to arbitrary excited states of one dimensional hermitian or nonhermitian systems. It has long been the objective to find a MER based formulation extendable to arbitrary bosonic and fermionic multidimensional states.  In this we have been succesfull, resulting in the second MER bounding formulation and the principal focus of the present work: the Orthogonal Polynomial Projection Quantization-Bounding Method (OPPQ-BM). Both EMM and OPPQ-BM exploit the importance of positivity/nonnegativity as a framework for quantization. EMM focuses on the positivity/nonnegativity properties of the  configuration space representation for the discrete state. OPPQ-BM focuses on the positivity of certain integral expressions. As such, the computational demands of OPPQ-BM are modest: only algebra is required. The computational demands of EMM are greater, requiring the use of convex optimization analysis. 

EMM primarily generates converging bounds; although it can also be viewed as an eigenenergy estimation method from the broader  perspective associated with an underlying Max-Min reformulation [8]. OPPQ-BM generates converging bounds as well, and high accuracy energy approximants that are consistent with these bounds. The EMM bounds are generated at any moment expansion order. The OPPQ-BM bounds require that we work at a sufficiently high moment expansion order, so that an upper bound to the limit of a certain converging positive sequence, can be approximated. Any crude upper bound to this limit can then be used to generate arbitrarily tight eigenenergy bounds. In this sense, EMM is more efficient than OPPQ-BM. To better appreciate this, we summarize below the principal contribution of the OPPQ-BM formalism, which is discussed in greater detail in Sec. 2 (i.e. for one dimensional problems) and Sec. 3 (i.e. for the multidimensional case). This is then followed by a more comprehensive discussion on the structure of EMM, so as to better contrast each approach with the other.

Given any MER quantum system,  OPPQ-BM generates a family of continuously differentiable, positive, increasing functions, on the energy axis: $0 < {\cal S}_i(E) < {\cal S}_{i+1}(E)$.  We note that for one space dimension configuration space problems, ${\cal S}_i(E) \equiv \lambda_i(E)$; whereas in multidimensions, ${\cal S}_i(E) \equiv {\cal L}_i(E)$, due to their distinct definitions. An additional important property   of these functions is that 
  in the infinite expansion limit, the ${\cal S}_i(E)$ become infinite almost everywhere except at the physical energies:
\begin{eqnarray}
Lim_{i\rightarrow \infty} {\cal S}_i(E) =\cases{ finite \iff E = E_{phys} \cr \infty \iff E \neq E_{phys} } .\nonumber
\end{eqnarray}
The local minima 
\begin{eqnarray}
\partial_E {\cal S}_i(E_i^{(min)})   = 0, \nonumber
\end{eqnarray}
  generate a positive increasing sequence which is bounded from above :
\begin{eqnarray}
0 < {\cal S}_i(E_i^{(min)})  <  {\cal S}_{i+1}(E_{i+1}^{(min)}) <  \ldots < {\cal S}_\infty(E_{phys}) = \ finite. \nonumber
\end{eqnarray}
The last relation underscores the fact that the physical energy are the ``local minima" for ${\cal S}_\infty(E)$.  It then follows that the local minima, at finite order, are the energy approximants to the corresponding physical energy:
\begin{eqnarray}
Lim_{i\rightarrow \infty} E_i^{(min)}  = E_{phys}. \nonumber
\end{eqnarray}  
Eigenenergy bounds are generated upon determining  any crude upper bound satisfying ${\cal B}_U > \{{\cal S}_i(E_i^{(min)})\}$.  If the sequence rapidly converges (which is the case for the problems considered here) then such upper bounds are easily determined. Knowledge of ${\cal B}_U$ allows us to generate arbitrarily tight eigenenergy bounds. This is because there will always be energy intervals satisfying $[E_i^{(L)},E_i^{(U)}] \equiv \{E|{\cal S}_i(E) \leq {\cal B}_U\}$, whose endpoints are the lower and upper bounds to the corresponding physical energy, $E_i^{(L)} < E_{phys} < E_i^{(U)}$,  converging to each other in the limit: $Lim_{i \rightarrow \infty}(E_i^{(U)}-E_i^{(L)}) = 0$. We emphasize that a crude upper bound ${\cal B}_U$ can generate arbitrarily tight eigenenergy bounds, provided one can generate the sequences to arbitrarily high order. Alternatively, a good ${\cal B}_U$ estimate can yield tighter bounds at lower expansion orders. The illustrative example of the quantum harmonic oscillator, particularly Figs 1-3 and Table 1, displays all the above features. 

We continue with a description of EMM, the other MER-Bounding formalism, to better contrast it with OPPQ-BM.

Through important theorems in mathematics, and convex optimization [9-11], intimately associated with the moment problem [12], control theory methods can be developed for achieving tight bounds on important physical parameters. For quantum operators of low dimension, the first efforts in this regard involved the Eigenvalue Moment Method (EMM) by Handy et al [5-7] which combined four important areas of mathematical and computational physics previously overlooked: (i) the existence of moment equation representations (MER); (ii) the positivity theorems for the bosonic ground state [13] (and the extension of positivity/nonnegativity considerations to excited states [14-16]); (iii) the moment problem (MP) positivity theorems [12]; and finally, (iv) advances in linear and nonlinear convex optimization [9-11]. 

The EMM method was used succesfully to generate converging bounds to the ground state binding energy for the notoriously difficult quadratic Zeeman (QZM) effect for strong-superstrong magnetic fields [6,7]. Indeed, the ground state is 
normally regarded as the most singular state of the system and the most difficult to determine. This SPSCS problem had been the focus of many different approximation methods, all yielding widely varying results, as reviewed in the work by LeGuillou and Zinn-Justin [17]. Their order dependent conformal analysis yielded approximants that were exceptionally consistent with the EMM generated bounds. This coincidence has not been explained, except for the important observation that EMM corresponds to an affine map invariant variational procedure that automatically samples over all affine map transformations of the polynomial trial functions considered. We refer to this particular EMM formalism as EMM-$\Psi$. 

The space of polynomial functions maps into itself under affine maps. Power moments then become the natural, extensive objects, with which to define a quantization strategy. This is the cornerstone of EMM. This theme reveals itself in other contexts as well. Thus, for iterated function system (IFS)  generated fractals [18], affine maps are an essential component, and MER type relations play a special role in addressing the inverse problem (i.e. given a fractal measure determine the affine maps that generated it) [19,20]. More recently, Klauder [21] has formulated an affine quantization formalism (where positivity plays an essential role) and which seems well suited for addressing problems that standard canonical quantization cannot solve. We believe that the MER-bounding procedures motivate the importance of a broader quantization framework.

The EMM procedure can be generalized to exploit the fact that excited states can usually be mapped into  representations in which they correspond to uniquely nonnegative and exponentially bounded configurations, when compared to unphysical solutions. Thus, for  any one dimensional potential, the probability density, $|\Psi(x)|^2$ will satisfy either a third order linear ordinary differential equation (LODE), if the potential is real [14], or a fourth order LODE, if the potential is complex [15,16]. If the potential is of rational fraction form, then these LODEs can be transformed into a MER representation for the power moments of the probability density, and EMM implemented, yielding bounds for all the discrete states including the complex eigenergies corresponding to pseudo-Hermitian systems. In this manner, the first accurate prediction for the onset of PT-symmetry breaking for the $V(x) = ix^3+iax$ potential was obtained [16], contradicting the asymptotic analysis of Delabaere and Trinh [22].  

Unfortunately, this type of analysis, referred to as  EMM-$|\Psi|^2$, cannot be extended to multidimensions. As such, the application of EMM-$\Psi$ is currently limited solely to the multidimensional bosonic ground state; although it does appear that a multidimensional extension of EMM-$\Psi$ to excited states can still be implemented by combining two prior results [8,23] with a quantization philosophy suggested by the present work. The results of this will be the focus of a forthcoming communication. 

It is important to note that the original EMM formulation involved a nonlinear convex optimization computational framework now referred to as semidefinite programming (SDP) [9,10]; however, this can be relaxed through linear convex optimization methods (i.e. linear programming (LP)), the computational basis for all EMM implementations [6,7]. A better statement is that nonlinear convex sets can be bounded by arbitraily tight, circumscribing, convex polytopes. The latter corresponds to the LP relaxation of the underlying nonlinear problem. EMM concerns itself primarily with the existence of such convex sets.

The proposed new method, OPPQ-BM, is partly based on a previous estimation method developed by Handy and Vrinceanu [24], referred to here as the Orthogonal Polynomial Projection Quantization-Approximation Method (OPPQ-AM).  Through the bounding capabilities of OPPQ-BM,  a more complete (theoretically and computationally) framework emerges when compared to its precursor, OPPQ-AM.  Despite this, the OPPQ-AM estimates can produce impressive results [24]. One particular example is the two dimensional quantum dipole (QD) problem analyzed by Amore and Fernandez [25] through their large order Rayleigh Ritz variational analysis,  and subsequently prosecuted by Handy and Vrinceanu through the  OPPQ-AM formalism [26]. The review of the literature by both groups of authors reveals the poor estimates generated by an assortment of alternate methods as developed by others. Despite this, the Amore and Fernandez upper bound results for the ground state are in slight disagreement with the OPPQ-AM results that generate converging estimates from below. Application of OPPQ-BM should resolve this discrepancy, the focus of a forthcoming work.

\subsection {The Moment Equation Representation}

Given the importance of MER representations,  we review its essential structure in anticipation of the more detailed analysis provided in the next two sections.

Many important, multidimensional, physical systems can be transformed into a moment equation representation (MER), in which the power moments of the discrete states satisfy a homogeneous, linear, finite difference equation, of (effective) order $1+m_s$.  These in turn lead to the recursive generation of the power moments through a  linear relation involving energy dependent coefficients, and a subset of initialization moment variables, referred to as the {\it missing moments}, symbolized by the $\mu_\ell$ notation:  
\begin{eqnarray}
\mu(p) = \sum_{\ell = 0}^{m_s} M_E(p,\ell) \mu_\ell ,\nonumber
\end{eqnarray}
for $p \geq 0$, where $\mu_\ell \equiv \mu(\ell)$, and $M_E(\ell_1,\ell_2) = \delta_{\ell_1,\ell_2}$, for $0 \leq \ell_{1,2} \leq m_s$.  These MER relations are obtained from the Schrodinger equation by projecting into the moment space through an integration by parts analysis specific to the discrete states. Since unphysical configuration space solutions to the Schrodinger equation have power moments that are infinite in magnitude, they are automatically filtered out by the very nature of the MER, since the missing moments only generate finite $\mu(p)$'s.

For one space dimension problems, the missing moment order is finite, $m_s < \infty$; whereas multidimensional problems involve infinitely many missing moments, $m_s = \infty$. Through the MER formalism, the entire set of moments, ${\cal U}_\infty \equiv \{\mu(p)| p \geq 0\}$, can be decomposed into an infinite hierarchy of finite dimensional subspaces within which the Schrodinger operator is projected exactly: ${\cal U}_\infty = \bigcup_{i=0}^\infty{\cal U}_i$, where $ {\cal U}_0 \subset {\cal U}_1 \subset \ldots \subset {\cal U}_i \subset \ldots \subset {\cal U}_\infty$.  The challenge is to identify quantization procedures, or constraints, that apply within each finite subspace. 

If the physical solution is uniquely associated with positive or nonnegative configurations [5,14-16] the well known theorems associated with the {\it Moment Problem} (MP) in mathematics [12] lend themselves to this procedure because the underlying Hankel-Hadamard determinantal positivity constraints can be applied just to those power moments within each finite subset, ${\cal U}_i$. These constraints in turn generate tight, converging lower and upper eigenenergy bounds on the discrete states associated with the positive or nonnegative discrete state configurations. This structure persists in multidimensions [6,7]. This type of analysis is the focus of EMM.

The MER formulation defines a higher order (dimension) space in which the kinetic energy appears as a regular perturbative term; therefore defining a better representation in which to study SCSPS type problems, such as QZM [6,7]. The latter continues to be an important model for developing new, high accuracy,
eigenenergy methods as recently investigated by Schimerczek and Wunner [27], and other authors cited in their work. 

The OPPQ-BM implementation for one space dimension problems is very different from the multidimensional case. The first is reviewed in Sec. 2; whereas the second is outlined in Sec. 3 and defines a new computational formalism in quantum physics: constrained quadratic form minimization (CQFM). 

We recall from the preceding discussion that the bounding structure of OPPQ-BM depends on determining the ${\cal B}_U$ upper bound. This upper bound is the result of assessing the behavior of all the ${\cal U}_i$ moment subspaces. For one space dimension problems, the ${\cal U}_i$ moment subspaces involve the same number of missing moments, provided $i \geq m_s$. As such, one can impose the same missing moment normalization across all these subspaces within OPPQ-BM. However, in the multidimensional case, as the expansion order of ${\cal U}_i$ increases, the number of missing moments that generate it, increases as well. Therefore, within OPPQ-BM great care is needed in defining a consistent normalization condition on the missing moments, across all the subspaces, simultaneously. The CQFM formalism does this exactly. By way of contast, the EMM bounds are generated within each ${\cal U}_i$ subspace (they are not dependent on the higher order subspaces); therefore EMM is not beset with the more complicated normalization requirements of OPPQ-BM. 

Illustrative examples of the above are afforded by the quantum harmonic oscillator in Sec.4, followed by the  application of OPPQ-BM to  the QZM problem in Sec. 5 (refer to Table 2). In the second case, we are able to match or surpass the high accuracy estimates  of Kravchenco et al [28] for a broad range of magnetic field strengths (i.e. $B \leq O(200)$, in atomic units) as applied to the ground  ($\epsilon_{gr}$) and first excited ($\epsilon_1$) states with  zero azimuthal angular momentum  and even parity (i.e. $0^+$), for simplicty. Specifically, we can generate OPPQ-BM estimates and bounds from 13-20 significant figures, except for $\epsilon_1$ at $B = 200$, where we can only attain five of the seven significant figures given by Kravchenko et al, within our current computing capabilities (i.e. MacBook Pro 2.2GHZ/1333MHz). For the  $0^+$ states, Kravchenco et al only provide results for  $\epsilon_{gr}$ for $B \leq 4000$, and $\epsilon_1$ for $B \leq 1000$. The OPPQ-BM results  for $\epsilon_{gr}$, at $B = 2000$,  agree to approximately nine significant figures. For higher magnetic field strengths (i.e. $B \leq O(10^4)$) we compare OPPQ-BM results to those of  Schimerczek and Wunner [27], who utilize a B-spline approach. Within the limits of our computing power,  the OPPQ-BM  ground state binding energy results surpass, or come close  to (i.e. seven of ten significant digits), their results. For $\epsilon_1$ at $ B = 2000$ and $B = 10^4$,  our computing power limitations  generate excited state energies of diminishing accuracy: $1\%$ and $14\%$ accuracy, respectively.   The present work involves no more than 50 variational parameters (i.e. $m_s \leq 50$ as indicated in Tables 2-5). On any faster computer platform than ours, one should be able to work with larger numbers of variational parameters permitting the accurate determination of these and other binding energies for larger magnetic field strengths. Also, there are alternative OPPQ-BM representations than that presented here which may lead to faster convergence at high magnetic field strengths. These are the subject of ongoing investigations. 

Despite the exceptionally impressive results by Kravchenco et al, their analysis is very intricate, and involves some truncations. Our OPPQ-BM analysis is straightforward with no truncations, and can produce tight bounds.   Faster results, with higher significant figures, can be readily obtained with the current formalism when implemented on faster platforms. The adopted algebraic software is Mathematica.
\newpage
\section{\bf The OPPQ-BM Formalism for 1-Space Dimension Problems}

Assume the system admits a MER relation for the power moments of the discrete states, $\mu(p) = \int dx \ x^p\ \Psi(x)$, satisfying the recursive structure
\begin{eqnarray}
\mu(p) = \sum_{\ell = 0}^{m_s} M_E(p,\ell) \mu_\ell, \ {\rm for}\ p \geq 0,
\end{eqnarray}
where $M_E(\ell_1,\ell_2) = \delta_{\ell_1,\ell_2}$ for $0 \leq \ell_{1,2} \leq m_s$. The MER relation defines a homogenous and linear relation amongst the moments. One can impose a normalization condition on the missing moments, for instance $\sum_{\ell = 0}^{m_s} \mu_\ell^2 = 1$.

Consider the expansion of the wavefunction, in terms of  the complete set of orthonormal polynomials with respects to a suitably chosen {\it reference \ function - weight}, $R(x)$:
\begin{eqnarray}
\Psi(x) = \sum_{i=0}^\infty c_i  P_i(x) R(x),
\end{eqnarray}
where 
\begin{eqnarray}
\langle P_i|R|P_j\rangle = \delta_{i,j},
\end{eqnarray}
and the orthonormal polynomial coefficients are represented by
\begin{eqnarray}
P_i(x) = \sum_{j=0}^i \Xi_j^{(i)} x^j.
\end{eqnarray}
We can then generate the $c_i$ expansion coefficients (implicitly for the discrete states) from the underlying MER relation: 
\begin{eqnarray}
\hspace{-35pt}c_i(E,\mu_\ell) = \langle P_i|\Psi\rangle  = \sum_{j=0}^i \Xi_j^{(i)} \mu(j) = \sum_{\ell = 0}^{m_s} \Lambda_{E,\ell}^{(i)} \ \mu_\ell, 
\end{eqnarray}
where
\begin{eqnarray}
\Lambda_{E,\ell}^{(i)} = \sum_{j=0}^i\Xi_j^{(i)} M_E(j,\ell). 
\end{eqnarray}
We emphasize that all the above relations are exact. No truncations are involved. 

The generation of the orthonormal polynomial coefficients is straightforward, requiring a Cholesky analysis.  It involves knowing the power moments of the weight, $w_i \equiv \int dx \ x^i R(x)$, and transforming Eq.(3) into $\langle {\overrightarrow \Xi}^{(I)}|W|{\overrightarrow \Xi}^{(J)}\rangle = \delta_{I,J}$, where $W_{i,j} = w(i+j)$ is the Hankel matrix of the weight function. If $W = C C^\dagger$ denotes the Cholesky decomposition, then ${\overrightarrow \Xi}^{(J)} = (C^{\dagger})^{-1}{\hat {\bf e}}^{(J)}$, where ${\hat{\bf e}}^{(J)}$ is the unit tuple in the $J$-th direction.

If the reference function decreases, asymptotically, no faster than the physical state, $\lim_{|x|\rightarrow \infty}|{{\Psi_{phys}(x)}\over{R(x)}}| \rightarrow const$, then the normalizability of the discrete states is given by
\begin{eqnarray}
\langle {{\Psi\over {\sqrt{R}}}}|{{\Psi\over {\sqrt{R}}}}\rangle = \langle \Psi|{1\over R}|\Psi \rangle = \sum_{i=0}^\infty c_i^2.
\end{eqnarray}
We can further relax the asymptotic conditions on the reference function provided the product $Lim_{|x|\rightarrow \infty} \Psi_{phys}(x) \times {{\Psi_{phys}(x)}\over{R(x)}} \rightarrow 0$, in a manner that makes the above integral finite, if $\Psi$ is a physical state, and infinite, if $\Psi$ is unphysical.

 Having chosen an appropriate reference function, Eq.(7) applied on the physical discrete states (i.e. physical energy and missing moment values) defines a convergent  positive series; and the high order coefficients must asymptotically vanish: $Lim_{i\rightarrow \infty}c_i = 0$. The latter is the quantization condition within OPPQ-AM (i.e. Handy and Vrinceanu [24,26]). For unphysical values of the energy and/or the missing moments, this power series must become infinite.

Let us represent the coefficients in terms of the normalized missing moment vectors (i.e. $\sum_{\ell = 0}^{m_s} \mu_\ell^2 = 1$):

\begin{equation}
c_i\big(E, \mu_0,\ldots,\mu_{m_s}\big) = {\overrightarrow \Lambda}^{(i)}_{E}\cdot {\hat \mu};
\end{equation}
and define the partial sums:
\begin{eqnarray}
\hspace{7pt} {\cal S}_I(E,{\hat \mu}) = \sum_{i=0}^I c_i^2, \\
\hspace{52pt} = \langle {\hat \mu} | {\cal P}_I(E) |{\hat \mu} \rangle,  \\
\hspace{17pt} {\cal P}_I(E) = \sum_{i=0}^I{\overrightarrow \Lambda}^{(i)}_{E} {\overrightarrow \Lambda}^{(i)}_{E},
\end{eqnarray}
involving an energy dependent, positive definite matrix (for $I \geq m_s$, since the $\Lambda$ vectors should be independent at this order). All these matrices are of the same dimension ($1+m_s$), if $I \geq m_s$. All the monotonic relations below stem from this uniform dimensionality.

The partial sums form a monotonically increasing, positive, sequence: 

\begin{eqnarray}
0  < {\cal S}_I(E,{\hat \mu}) < {\cal S}_{I+1}(E,{\hat \mu}) < \ldots .
\end{eqnarray}
In terms of the partial sums, the OPPQ-BM quantization conditions become

\begin{eqnarray}
Lim_{I\rightarrow \infty}{\cal S}_I(E,{\hat \mu})  = \cases{  \rm finite,\ only\ for\  E = E_{phys}\ {\underbar {and}}\ {\hat \mu} = {\hat \mu}_{phys};\cr
\infty,\ if\  E \neq E_{phys}\ {\underbar {or}}\ {\hat \mu} \neq {\hat \mu}_{phys},\ {\underbar {or}}\ both .}
\end{eqnarray} 
The principal focus of OPPQ-BM is devising an effective strategy for identifying the physical solutions satisfying Eq.(13).  To this extent, we can argue the following:\\
\\
\hspace{-22pt}{\textbf {Theorem 1}:}
\begin{equation}
{\hat \mu}_{phys} = Eigenvector\ of\ the\ Smallest\ Eigenvalue\ of\ {\cal P}_\infty(E_{phys}).\\
\end{equation}
The proof is straightforward since if this is not the case then it would contradict Eq.(13), which is a consequence of the finiteness of Eq.(7) for physical states. Let  $\lambda_*$ be the smallest eigenvalue  for the positive definite, symmetric, matrix ${\cal P}_{\infty}(E_{phys})$. Let ${\hat \mu}_{phys} $ be the normalized vector that satisfies $\langle {\hat \mu}_{phys} |{\cal P}_{\infty}(E_{phys})| {\hat \mu}_{phys}\rangle < \infty$.  Since $\lambda_* \leq  \langle {\hat \mu}_{phys} |{\cal P}_{\infty}(E_{phys})| {\hat \mu}_{phys}\rangle < \infty$, if  $\lambda_*$ does not correspond to a physical state, then $\lambda_* = \infty$, resulting in a contradiction. Therefore, Theorem 1 is established. 

Let us now define the smallest eigenvalue for the positive, symmetric matrix ${\cal P}_I(E)$:
\begin{equation}
\lambda_I(E)   \equiv Smallest\ Eigenvalue\ of\ {\cal P}_I(E).
\end{equation}
It then follows that these also form (essentially) a monotonically increasing sequence:
\begin{equation}
0 < \ldots < \lambda_I(E)   < \lambda_{I+1}(E) < \ldots <\lambda_{\infty}(E).
\end{equation}
This is because ${\cal P}_{I+1}(E) = {\cal P}_I(E) + {\overrightarrow \Lambda}_{E}^{(I+1)} {\overrightarrow \Lambda}_{E}^{(I+1)}$; and for any vector, ${\overrightarrow A}$, we must have $\langle {\overrightarrow A}|{\cal P}_{I+1}(E) |{\overrightarrow A}\rangle = \langle {\overrightarrow A}|{\cal P}_{I}(E) |{\overrightarrow A}\rangle + \langle {\overrightarrow A}|{\overrightarrow \Lambda}_{E}^{(I+1)} {\overrightarrow \Lambda}_{E}^{(I+1)} |{\overrightarrow A}\rangle$ or $\langle {\overrightarrow A}|{\cal P}_{I+1}(E) |{\overrightarrow A}\rangle > \langle {\overrightarrow A}|{\cal P}_{I}(E) |{\overrightarrow A}\rangle$. This is a strict inequality unless ${\overrightarrow A}$ is the null vector of the dyad term, or $\langle {\overrightarrow \Lambda_E}^{(I+1)}|{\overrightarrow A}\rangle = 0$. For a fixed, arbitrary $E$, such occurences as $I \rightarrow \infty$ are expected to be very rare given that the dimensionality of the vectors is $1+m_s$.

From the OPPQ-BM quantization condition in Eq.(13), we then must have:
\begin{eqnarray}
Lim_{I \rightarrow \infty} \lambda_I(E) = \cases{ finite, {\rm \ for\ E =\ E_{phys};} \cr
\hspace{20pt}\infty, \hspace{4pt} {\rm for\ E \neq\ E_{phys}} .\cr}
\end{eqnarray} 

Clearly, the impact of Eq.(17) is that in a neighborhood of a physical energy, ${\cal N}(E_{phys})$, the locally concaved (upwards) eigenvalue functions, $\lambda_I(E)$, will nest above each other, with increasing index, $I$. The local minima for each (both in terms of location along the energy axis, and function value) must converge, generating in turn a sequence of converging eigenenergy approximants to the physical value, $E_{phys}$. The illustration in Fig.1 for the harmonic oscillator manifests the general behavior of these eigenvalue functions, for any one dimensional system. To validate this, define the local minima for each eigenvalue function of order $I$:

\begin{eqnarray}
\partial_E\lambda_I(E_I^{(min)}) = 0, \ for \ E_I^{(min)} \in {\cal N}(E_{phys}).
\end{eqnarray}
Since $ \lambda_{I}(E_{I}^{(min)}) < \lambda_{I}(E_{I+1}^{(min)}) < \lambda_{I+1}(E_{I+1}^{(min)})$, these matrix eigenvalues must converge from below to $\lambda_\infty(E_{phys})$. This result becomes: \\
\\
\noindent {\bf{Theorem 2}}: Within a neighborhood of a physical energy value, ${\cal N}(E_{phys})$, the local minima of the $\lambda_I(E)$ functions, $\partial_E\lambda_I(E_I^{(min)}) = 0$, generate a convergent, (essentially) monotonically increasing, positive, sequence of matrix eigenvalues converging from below to $\lambda_\infty(E_{phys})$.
\begin{eqnarray}
\lambda_{I}(E_{I}^{(min)}) < \lambda_{I+1}(E_{I+1}^{(min)}) < \ldots  <\lambda_\infty(E_{phys}) < \infty.
\end{eqnarray}

\noindent {\bf{Theorem 3}}: 
\begin{eqnarray}
Lim_{I\rightarrow \infty} E_{I}^{(min)} = E_{phys}.
\end{eqnarray}
The $\lambda$-eigenvalues converge monotonically; however, the local minima in the energy variable will converge to $E_{phys}$ but not necessarily monotonically. 

All the above lead to the bounding procedure summarized in Theorem 4.
\\

\noindent {\bf{Theorem 4:}} {{{The OPPQ-BM Eigenenergy Bounding Procedure}, for one dimensional systems}:

Let ${\cal B}_U$ be any (empirically determined) upper bound to the limit $\lambda_\infty(E_{phys})$,
\begin{equation}
{\cal B}_U> \lambda_\infty(E_{phys} ),
\end{equation}
 then it will generate arbitrary tight bounds to the physical energy. 

Theorem 4  is due to the fact that the $\lambda_I(E)$ are concaved upward with one minimum in ${\cal N}(E_{phys})$; and so long as  $E$ is close to (but not equal to) the physical energy, $E_{phys}$, we must have $Lim_{I\rightarrow \infty}\lambda_I(E) = \infty$. There then will be two roots satisfying:
\begin{eqnarray}
\lambda_I(E_I^{(L)}) = {\cal B}_U, \ and\ \lambda_I(E_I^{(U)}) = {\cal B}_U,
\end{eqnarray} 
which must converge to a point, in the infinite limit:
\begin{equation}
Lim_{I\rightarrow \infty} \Big( E_I^{(U)}- E_I^{(L)}\Big) = 0,
\end{equation}
and bound the physical energy:
\begin{equation}
E_I^{(L)} < E_{phys} < E_I^{(U)}.
\end{equation}

The upper bound ${\cal B}_U$ is empirically determined, once the convergence of the sequence in Eq.(19) is established. Clearly, the closer ${\cal B}_U$ can be reliably approximated to $\lambda_\infty(E_{phys})$, the tighter will be the bounds at lower $I$-index values. 
\newpage
\section{OPPQ-Bounding Method : Extension to Multidimensions}

In preparation for the quadratic Zeeman (QZM) problem, assume that $\Psi(x,y)$ denotes the wavefunction on the nonnegative quadrant, $\Re^2$. The power moments, $\mu({\overrightarrow p})= \int dx \int dy\ x^{p_1}y^{p_2}\Psi(x,y)$, are generated through a moment equation recursive structure of the form
\begin{equation}
\mu({\overrightarrow p}) =\sum_{\ell=0}^{m_s} M_E({\overrightarrow p},\ell) \ \mu_{\ell}, \ {\overrightarrow p} \in {\cal P}_{m_s}, \ m_s = 0, 1,\ldots \infty,
\end{equation}
where $\mu_\ell \equiv \mu(p_{1;\ell},p_{2;\ell})$, are a subset of the moments.
That is, at any one time one is working with a finite number of two dimensional moments, generated by a corresponding finite number of missing moments. This structure persists in a sequential manner. Thus, 
the first $1+m_s$ missing moments will generate all the power moments in the finite set ${\cal U}_{m_s}$, etc.  These sets form an infinite hierarchy of moment subspaces: ${\cal U}_{m_s} \subset {\cal U}_{m_s+1} \subset \ldots \subset {\cal U}_\infty$. 

A normalization condition on the infinitely many missing moments is required. The OPPQ-BM philosophy requires that we relate the missing moments in ${\cal U}_{m_s}$ to the (increasing number of) missing moments in ${\cal U}_{m_s+1}$, etc.  For one space dimension problems, since the number of missing moments is (essentially) constant across all moment subspaces, a uniform normalization across all the corresponding ${\cal U}$ subspaces is possible; and OPPQ-BM can be implemented. This is not the case for problems with spatial dimension two, or greater.  It is also highly unlikely, that in the infinite limit, imposing the normalization $\sum_{\ell = 0}^\infty \mu_\ell^2 = 1$, is physically possible, since this would imply that the higher order missing moments go to zero, asymptotically. Alternatively, one can impose a normalization condition involving only a finite number of the missing moments, across all subspaces. The simplest possibility (not expected to violate any symmetry conditions for the QZM states of interest) would be taking

\begin{eqnarray}
\mu_0 = 1,
\end{eqnarray}
the choice adopted in this work. Accordingly, once a normalization is adopted, our OPPQ-BM analysis will restrict itself to missing moment solutions satisfying the adopted normalization.

The two dimensional OPPQ-BM formalism is identical to the one dimensional case. We summarize this here. Thus, if $R(x,y)$ denotes the adopted weight, then $\Psi(x,y) =\sum_{i=0}^\infty c_i  P_i(x,y) R(x,y)$ is the orthonormal polynomial decomposition of the discrete states, based on some chosen sequential ordering of the orthonormal polynomials. We can write $P_i(x,y) = \sum_{j=0}^i \Xi_j^{(i)} x^{m_j}y^{n_j}$, with  $c_i = \langle P_i|\Psi\rangle =\sum_{j=0}^i \Xi_j^{(i)} \mu({m_j},{n_j})$ or $c_i = \sum_{j=0}^i \Xi_j^{(i)} \sum_{\ell=0}^{m_s(i)} M_E({m_j},{n_j},\ell) \mu_\ell$, where $m_s(i)$ is the maximum missing moment order required to generate the first $1+i$ projection coefficients. Let $I_{m_s}$ correspond to the first $1+I_{m_s}$ projection coefficients (i.e. $c_i$ where $0 \leq i \leq I_{m_s}$) that can be generated from all the $1+m_s$ missing moments $\{\mu_\ell, 0 \leq \ell \leq m_s\}$. That is, we assume that the coordinate pair ordering $(m_i,n_i)$, and the missing moment structure, has been chosen to allow for a sequential progression of $m_s$ and $I_{m_s}$ (i.e. $I_{m_s+1} > I_{m_s}$). We then have $c_i(E,\mu_0,\ldots, \mu_{m_s}) = \sum_{\ell = 0}^{m_s} \Lambda_{E,\ell}^{(i)} \mu_\ell$, where $\Lambda_{E,\ell}^{(i)} = \sum_{j=0}^i \Xi_j^{(i)} M_E({m_j},{n_j},\ell)$.

We define the partial sums:

\begin{eqnarray}
{\cal S}_{I_{m_s}}(E,\mu_0,\ldots, \mu_{m_s}) = \sum_{i=0}^{I_{m_s} }c_i^2, \\ \cr
\hspace{108pt} = \langle {\overrightarrow \mu}|\sum_{i=0}^{I_{m_s} } \overrightarrow{\Lambda^{(i)}_{E}}\overrightarrow{\Lambda^{(i)}_{E}}|{\overrightarrow \mu}\rangle \equiv \langle {\overrightarrow \mu}|{\cal P}_{I_{m_s}}(E)|{\overrightarrow \mu}\rangle. 
\end{eqnarray}
These expectation values involve  ${\cal P}_{I_{m_s}}$,  a positive definite matrix of dimension $(m_s +1)\times (m_s+1)$; and are dependent on the $\{\mu_\ell|0 \leq \ell \leq m_s\}$ missing moments that generate the ${\cal U}_{m_s}$ moment subspace.  Due to their increasing dimensionality, the eigenvalues of these matrices will not necessarily satisfy the monotonically increasing property of their 1-space dimension counterpart in Eq.(16). However, for an arbitrary missing moment vector, ${\overrightarrow \mu}$, we have the monotonically increasing property $\langle {\overrightarrow \mu}|{\cal P}_{I_{m_s}}(E)|{\overrightarrow \mu}\rangle \leq \langle {\overrightarrow \mu}|{\cal P}_{I_{m_s+1}}(E)|{\overrightarrow \mu}\rangle$, or

\begin{equation}
{\cal S}_{I_{m_s}}(E,\mu_0,\ldots, \mu_{m_s}) < {\cal S}_{I_{m_s+1}}(E,\mu_0,\ldots, \mu_{m_s},\mu_{m_s+1}).\\
\end{equation}
These must satisfy the multidimensional OPPQ-BM quantization conditions:

\begin{eqnarray}
\hspace{-40pt} Lim_{m_s \rightarrow \infty} {\cal S}_{I_{m_s}}(E,\overrightarrow{\mu}) =  \cases{
  {finite, \ if\ E = E_{phys}\ {\underbar {and}} {\overrightarrow \mu} = {\overrightarrow \mu_{phys}}},  \cr
{\infty, if \ E \neq E_{phys}\ {\underbar {or}}\ {\overrightarrow \mu} \neq {\overrightarrow \mu_{phys}}},} 
\end{eqnarray}
together with the adopted normalization constraint $\mu_0 = 1$.
\\

Given all the above, combined with the normalization condition in Eq. (26), the natural choice in implementing OPPQ-BM (i.e. satisfying Eq.(30)) for problems of spatial dimension greater than unity is to focus on the constrained optimization problem defined below.

\subsection{\textbf {Constrained Quadratic Form Minimization}}

Consider the constrained quadratic form generated by ${\cal P}_{I_{m_s}} \equiv D(E)$ in Eq.(28), incorporating the normalization condition $\mu_0 = 1$:

\begin{eqnarray}
\hspace{-65pt} \langle {\overrightarrow \mu}|D(E) |{\overrightarrow \mu}\rangle \equiv (D(E))_{0,0}+ 
2 \sum_{\ell=1}^{m_s}(D(E))_{0,\ell}\ \mu_\ell + \sum_{\ell_1=1}^{m_s}\sum_{\ell_2 = 1}^{m_s} \mu_{\ell_1}\ (D(E))_{\ell_1,\ell_2}\ \mu_{\ell_2},
\end{eqnarray}
or
\begin{eqnarray}
\langle {\overrightarrow \mu}|D(E) |{\overrightarrow \mu}\rangle \equiv C(E)+
 2 {\overrightarrow B}(E) \cdot {\overrightarrow u} + \langle {\overrightarrow u}|{{\textbf A}(E)}|{\overrightarrow u}\rangle,
\end{eqnarray}
where ${\overrightarrow u} \equiv (\mu_1,\ldots,\mu_{m_s})$, ${\overrightarrow \mu} = (1,{\overrightarrow u})$, and the other expressions implicitly defined. Then the constrained (global) minimization, in the unconstrained missing moment space, is satisfied by
\begin{eqnarray}
{\overrightarrow u}_{opt. soln} \equiv {\overrightarrow u}_0 = -{\textbf A}^{-1}(E)\ {\overrightarrow B}(E),
\end{eqnarray}
yielding the global minimimum for the constrained quadratic form:
\\
\begin{eqnarray}
\hspace{-60pt} {\cal L}_{I_{m_s}}(E) \equiv Min_{{\overrightarrow u}}\Big(\langle {\overrightarrow \mu}|D(E) |{\overrightarrow \mu}\rangle\Big) =C(E) -\langle {\overrightarrow B}(E) |{{\textbf A}^{-1}(E)}|{\overrightarrow B}(E) \rangle.
\end{eqnarray}

That is, Eq.(33) becomes the counterpart to Eq.(15), or identifying the smallest eigenvalue eigenvector in the one dimensional case. The counterpart to the energy dependent eigenvalue for the one dimensional case becomes the expression in Eq.(34) which defines the value of the quadratic form at the local minimum, ${\cal L}_{I_{m_s}}(E)$. 

From Eq.(29) it follows that these expressions generate the desired monotonically increasing sequences in the energy variable:

\begin{eqnarray}
0 < {\cal L}_{I_{m_s}}(E) < {\cal L}_{I_{m_s+1}}(E)  <  {\cal L}_{I_{m_s+2}}(E)\ldots.
\end{eqnarray}

For the adopted normalization $\mu_0 = 1$, these energy expressions become the multidimensional counterpart to the one dimensional $\lambda_I(E)$ monotonic relations in Eq.(16). From Eq.(30), the OPPQ-BM quantization conditions, we can only conclude that 
in the infinite (missing moment) expansion limit, $I_{m_s} \rightarrow \infty$, the local minima in the  energy variable define the physical values:

\begin{eqnarray}
{\cal L}_{\infty}(E) = \cases{  {finite, \iff \ E = E_{phys}\  },  \cr {\infty, \iff \ E \neq E_{phys} }.} 
\end{eqnarray}

The local minima in the energy variable, 
\begin{eqnarray}
\partial_E{\cal L}_{I_{m_s}}(E_{I_{m_s}}^{(min)}) = 0,
\end{eqnarray}
become the physical energy approximants (i.e. the counterpart to the 1-space dimension expression: $\partial_E\lambda_I(E_I^{(min)}) = 0$). 
It then follows that 

\begin{eqnarray}
\hspace{-50pt} {\cal L}_{I_{m_s-1}}(E^{(min)}_{I_{m_s-1}})  < {\cal L}_{I_{m_s}}(E_{I_{m_s}}^{(min)}) < {\cal L}_{I_{m_s+1}}
(E_{I_{m_s+1}}^{(min)}) < \ldots < {\cal L}_{\infty}(E_{phys}) = finite, \cr
\end{eqnarray}
and we can generate converging bounds in a manner identical to that for the 1-space dimension problem.

We note that for the one space dimension problem, and the multidimensional problem, the corresponding derivative expressions $\partial_E\lambda_I(E)$ and $\partial_E {\cal L}_{I_{m_s}}(E)$, respectively, can be generated in closed form through recursive relations originating within the MER formulations. The details are not given here; however being able to generate these expressions in closed form facilitates the use of bisection methods for determining the precise location of the respective minima in the energy variable. The results in Tables 1-5 make use of these expressions. 
\newpage
\section {The Quantum Harmonic Oscillator}

The quantum harmonic oscillator illustrates the main results of the previous analysis which applies to any one dimensional system.
For simplicity, we restrict our analysis to the even parity states. The same can be formulated with respects to the odd parity states, or alternatively unified to include both states.

For the harmonic oscillator, $-\partial_x^2 \Psi(x) + x^2\Psi(x) = E \Psi(x)$, the even parity states  satisfy a moment equation given by (i.e. multiply both sides by $x^{2p}$ and integrate by parts)
\begin{equation}
u(p+1) = E\ u(p)+2p(2p-1)\ u(p-1),
\end{equation}
$\ p \geq 0$.
These correspond to the Stieltjes moments $u(p) = \int_0^\infty d\xi \ \xi^p {{\Psi(\sqrt{\xi})}\over {\sqrt{\xi}}}$. 

Accordingly,
\begin{equation}
u(p) = M_E(p,0) u(0),
\end{equation}
$ p \geq 0$, where $M_E(0,0) = 1$ and $M_E(p,0)$ satisfies the moment equation with respect to the $p$-index. We note that we can take $u(0) =1$ for all even parity states.

The $M_E(p,0)$ coefficients satisfy the moment equation with respect to the $p$-index, subject to the initialization condition $M_E(0,0) =1$:
\begin{equation}
M_E(p+1,0) = E\ M_E(p,0)+2p(2p-1)\ M_E(p-1,0),\ p \geq 0.
\end{equation}
These correspond to polynomials in the energy parameter, $E$.
We can generate the orthonormal polynomials with respect to the weight $R(x) = e^{-{{x^2}\over 2}}$.

The even order orthonormal polynomials in the `$x$' variable generate the regular orthonormal polynomials in the `$\xi$' variable, $P_{n=even}(x) \equiv {\cal P}^{(\eta = n/2)}(\xi)$, where : 
\begin{eqnarray}
{\cal P}^{(I)}(\xi) = 
  {{(-{1\over 2})^I}
} {{( (2I)!)^{1\over 2}}\over{(2\pi)^{1\over 4}}} \sum_{i=0}^I{{(-2)^i}\over{(I-i)!(2i)!}}\xi^i .
\end{eqnarray}
The latter are orthonormal relative to the weight ${{exp(-\xi/2)}\over\sqrt{\xi}}$. The orthonormal polynomial coefficients for ${\cal P}^{(I)}(\xi) = \sum_{i=0}^I \Xi_i^{(I)}\xi^i$, are given by Eq.(42) . We then obtain the $c_i$ coefficients  (i.e. Eqs.(5-6))
\begin{equation}
c_i(E) = \sum_{j=0}^i\Xi_j^{(i)}M_E(j,0), i \geq 0.
\end{equation}
One can show that $c_i(E) = N_i \Pi_{j=1}^i (E-(1+4(j-1)))$, the roots being the exact even parity state energy values. Thus the OPPQ-AM (i.e. Handy and Vrinceanu [24,26]) quantization condition (i.e. $Lim_{i\rightarrow \infty} c_i(E) = 0$) is trivially satisfied.\\
\newpage
\begin{figure}
\includegraphics{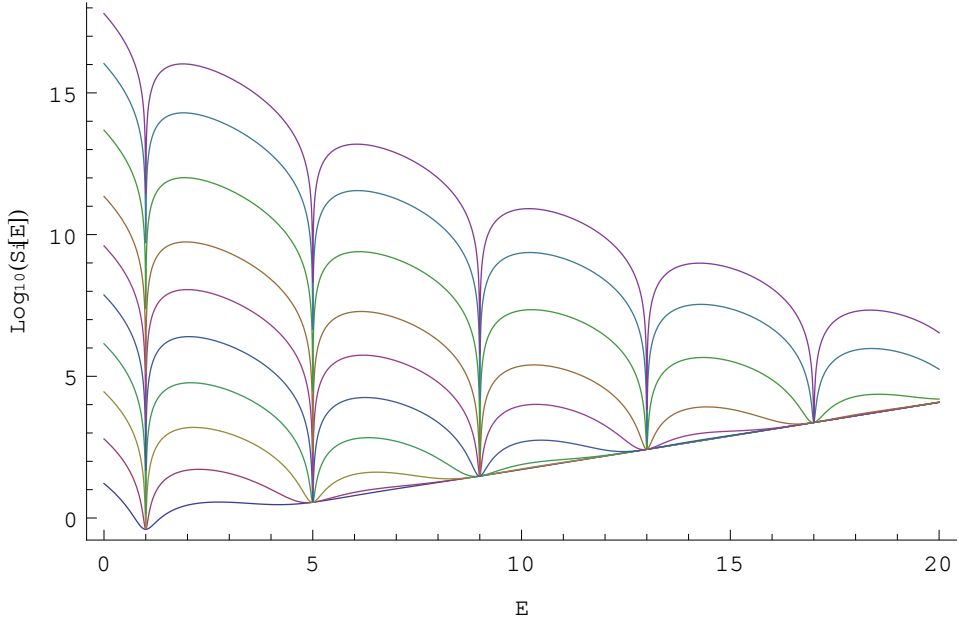}
\centering{\caption{ $Log_{10}(S_I(E))$ for even parity  states of the harmonic oscillator (i.e. $S_I(E) \equiv \lambda_I(E)$); $I = 5,8,11,\ldots, 32$.}}
\end{figure}
\newpage

In Fig. 1, we plot $Log_{10} \Big( \lambda_I(E)\Big)$ (i.e. $S_I(E) \equiv \lambda_I(E)$) over interval $0 \leq E \leq 20$.  The nesting of the $\lambda_I(E)$ curves is readily apparent. Although these functions are nested within each other, their local minima do not necessarily coincide (i.e. as clearly shown in Fig.3 for the second excited state). 

In Fig. 2  we show the progression of localized concavity around the ground state energy ($E_{gr} = 1$)  for the lower order partial sums, $\{\lambda_I(E)| 3 \leq I \leq 12\}$.  Indeed, the $\lambda_I(1)$ sequence is $\lambda_0(1) = \lambda_1(1) = \ldots = {1\over{\sqrt{2\pi}}} = .398942$; thereby concluding, within our OPPQ-BM formulation, that the ground state energy is precisely 1. 

\begin{figure}
\includegraphics{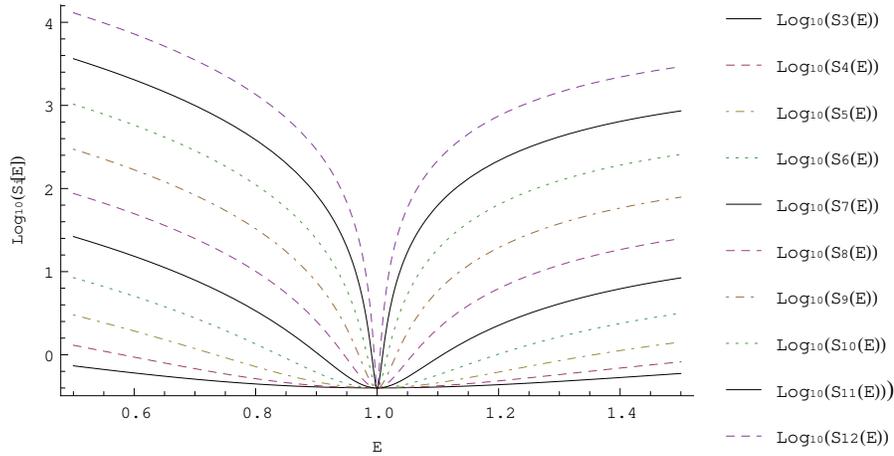}
\centering{\caption{Nesting of the partial sums $Log_{10}(S_I(E))$ centered around the ground state (i.e. $E_{gr} = 1$) for the harmonic oscillator, where $I = 3,4,5,\ldots, 12$. Note that all curves share the same, fixed, minimum.}}
\end{figure}

Things are more interesting for the second excited state, as given in Fig. 3 and  Table I 
. We determine the local minima for $\partial_E\lambda_I(E_I^{(min)}) = 0$, and generate the sequence $\{\lambda_I(E_I^{(min)})\}$ whose convergence  defines ${\lambda}_\infty(E_2) = 3.5904802$. A coarse upper bound ${\cal B}_U = 3.6 > {\lambda}_\infty(E_2) $ then allows us to generate converging bounds to the excited state by taking $\lambda_I(E_I^{(L)}) = \lambda_I(E_I^{(U)}) = {\cal  B}_U$, for $I \rightarrow \infty$. Note that the coarseness of the upper bound estimate for ${\cal B}_U$ does not determine the tightness of the eigenenergy bounds (which depend only on the expansion order $I$).

\begin{figure}
\includegraphics{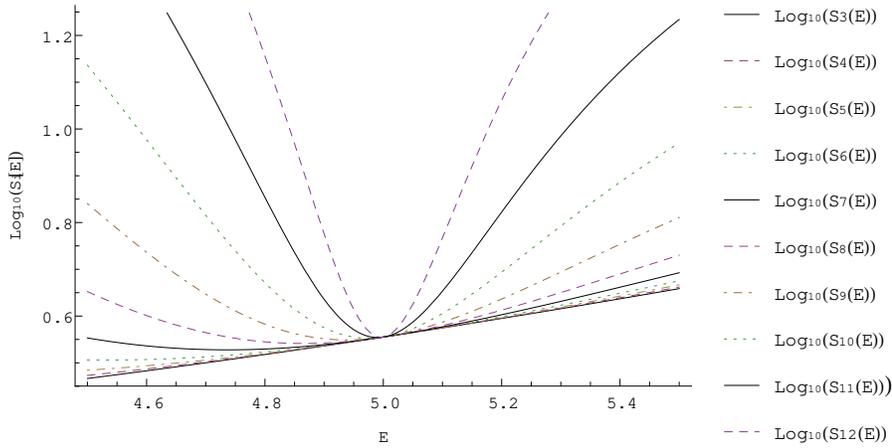}
\centering{\caption{Nesting of the partial sums $Log_{10}(S_I(E))$ (i.e. $S_I(E) \equiv \lambda_I(E)$) centered around the $2^{nd}$ excited state, $E_2 = 5$, for the harmonic oscillator, where $I = 3,4,5,\ldots, 12$. Note that their respective minima, in the energy variable,  monotonically increase to $E= 5$; and they all have the same derivative at that point.}}
\end{figure}

\begin{table}
\caption{
OPPQ-BM  for $E_2$:  $V(x) = x^2$, $R=e^{-{{x^2}\over2}}$ (i.e. $\lambda_I(E) \equiv S_I(E)$)
}
\centerline{
\begin{tabular}{lcccc}
\hline
$I$ & $\partial_E\lambda_I(E_I^{(min)})=0$    &  $\lambda_I( E_I^{(min)})$  & $E_I^{(L)}$     &  $E_I^{(U)}$     \\
\hline
\hline
 6 & 4.53222&  3.20587&  4.07088&  5.00593\\
 7 & 4.73661& 3.37132&  4.48590&  5.00591 \\
 8 & 4.86462&  3.47875&  4.73214&  5.00585\\
 9 & 4.93802&  3.54002&  4.87312& 5.00572 \\
 10 & 4.97454&  3.56996&  4.94437& 5.00541 \\
 11 & 4.99037&  3.58276&  4.97612& 5.00479 \\
 12 & 4.99656&  3.58773&  4.98933&  5.00384 \\
 13 & 4.99882&  3.58954&  4.99489&5.00276   \\
 14 & 4.99961&  3.59017&   4.99741& 5.00181  \\
\hline
20 &   4.9999996 &3.5904802 &4.99993 & 5.00007\\
\hline
 $\infty$ & 5&  $3.5904805< {\cal B}_U = 3.6$ &  5& 5 \\
 \hline
\end{tabular}}
\end{table}

\section{The Quadratic Zeeman Problem}

  For simplicity, we examine  the even parity, zero azimuthal angular momentum states, for the quadratic Zeeman (QZM) problem corresponding to: 
\begin{equation}
\Big( - {1\over 2} \Delta + {{B^2}\over 8} (x^2 + y^2) - {Z\over r} - E \Big ) \Psi = 0.
\end{equation}
We adopt the parabolic coordinate representation formalism used by Handy et al [7], transforming the three dimensional QZM problem  (atomic units adopted),
into a parabolic coordinate representation defined by $\xi = r-z \geq 0$, $\eta = r+z \geq 0$. Additionally, from EMM we know that a more efficient missing moment structure is obtained if we transform the wavefunction according to 
\begin{equation} 
\Phi(\xi,\eta) \equiv \Psi(\xi,\eta) exp(-B \xi \eta/4).
\end{equation}
The transformed parabolic partial differential equation becomes
\begin{eqnarray}
\hspace{-50pt} \partial_\xi(\xi\partial_\xi \Phi) + \partial_\eta(\eta\partial_\eta \Phi) + {1\over 2} B\xi\eta(\partial_\xi\Phi+\partial_\eta\Phi) 
+\Big[ {1\over 2}(E + {1\over 2} B) (\xi+\eta) + 1 \Big] \Phi = 0.
\end{eqnarray}
The asymptotic form of the transformed configuration is given by 
\begin{equation}
\Phi(\xi,\eta) \rightarrow exp\Big[ -{1\over 2}B\xi \eta -({\epsilon\over 2})^{1\over 2}|\eta-\xi|\Big],
\end{equation}
where the binding energy is given by $\epsilon = B/2 - E$.

The two dimensional Stieltjes moments for $\Phi$ are defined by
\begin{equation}
u(m,n) = \int_0^\infty d\xi \ \int_0^\infty d\eta \ \xi^m \eta^n \Phi(\xi,\eta),
\end{equation}
with moment equation
\begin{eqnarray}
\hspace{-50pt} m^2u(m-1,n)+n^2u(m,n-1) \cr 
-{1\over 2}[Bm+\epsilon]u(m,n+1) 
-{1\over 2}[Bn+\epsilon]u(m+1,n)+Zu(m,n) = 0,
\end{eqnarray}
with even parity invariance ($z \leftrightarrow  -z$ or $\xi \leftrightarrow \eta$) reflected in the moment reflection symmetry $u(m,n) = u(n,m)$. 

The moment equation defines a ``nearest neighbor" pattern in which the $u(m,n)$ moment is linked to the $\{u(m+1,n),u(m-1,n),u(m,n+1),u(m,n-1)\}$ moments, so long as the reflection symmetry is exploited, and the moment indices limited to the nonnegative integers $m,n \geq 0$. 
The missing moments correspond to  $\{u(\ell,\ell) | \ell \geq 0\}$.  For $0 \leq \ell \leq m_s$, the $1+m_s$ missing moments, $u(\ell,\ell) \equiv u_\ell$,  generate all the moments defined through their antidiagonal index: $\{u(m,n)| m+n \leq 2m_s+1\}$. In this manner we generate the moment - missing moment relation:
\begin{equation}
u(m,n) = \sum_{\ell = 0}^{m_s} M_\epsilon(m,n,\ell) u_\ell, \ where \ 0\leq m+n \leq 2m_s+1,
\end{equation}
$u_\ell \equiv u(\ell,\ell)$ and $M_\epsilon(\ell_1,\ell_1,\ell_2) = \delta_{\ell_1,\ell_2}$.  For a given $m_s$, only these moments can be generated. 

The binding energy matrix coefficients, $M_\epsilon(m,n,\ell)$, satisfy the moment equation with respect to the $(m,n)$ indices and the given initialization conditions:

\begin{eqnarray}
\hspace{-70pt} m^2M_\epsilon(m-1,n,\ell)+n^2M_\epsilon(m,n-1,\ell)  -{1\over 2}[Bm+\epsilon]M_\epsilon(m,n+1,\ell) \cr 
\hspace{70pt} -{1\over 2}[Bn+\epsilon]M_\epsilon(m+1,n,\ell)+ZM_\epsilon(m,n,\ell) = 0,
\end{eqnarray}
where $M_\epsilon(\ell_1,\ell_1,\ell_2) = \delta_{\ell_1,\ell_2}$.

The preferred  {\it reference function - weight} is any expression which takes on the  asymptotic form of the physical solutions. Instead of using the expression in Eq.(47), an easier expression to use (with respect to generating the required power moments of the weight, as required for the Cholesky generation of the orthonormal polynomials) is

\begin{equation}
R_{QZM}(\xi,\eta) = exp\Big( -{1\over 2}B\xi\eta - ({\epsilon\over 2})^{1\over 2} (\xi+\eta)\Big),
\end{equation}
with  power moments 
\begin{equation}
w_{QZM}(m,n) =  \int_0^\infty d\xi \int_{0}^\infty d\eta \  \xi^m \eta^n  exp\Big( -\beta\xi\eta - \alpha (\xi+\eta)\Big),  \nonumber 
\end{equation}
\begin{equation}
\hspace{65pt} \equiv {{n!}\over{\alpha^{m+n+2}}}\Gamma(m,n+1,g)
\end{equation}
where $\alpha = ({\epsilon\over 2})^{1\over 2} $, $\beta = {1\over 2}B$, and $g = {\beta \over{\alpha^2}} = {B\over \epsilon}$. The $\Gamma$ functions are recursively generated as follows. First, $\Gamma(0,1,g) < 1$ is numerically determined to high accuracy. This then allows us to generate
\begin{equation}
\Gamma(0,n+1,g) = \sum_{j=1}^n {{(-1)^{j+1}}\over {g^j}}{{(n-j)!}\over{n!}}  +{{(-1)^{n}}\over{g^nn!}}\Gamma(0,1,g),
\end{equation}
for $n \leq N$. For each such `$n$', we can generate 
\begin{equation}
\hspace{-70pt} \Gamma(m+1,n+1,g) = {1\over g} \delta_{m,0} + {m\over g} \Gamma(m-1,n+1,g) + [m-n-g^{-1}] \Gamma(m,n+1,g), 
\end{equation} 
for $0 \leq m \leq M$.

One can allow the reference function to incorporate the binding energy parameter, as given above. This makes the generation of the orthonormal polynomials more time consuming. We do this to low order to obtain an estimate of the physical binding energy (i.e. $\epsilon_0 \approx \epsilon_{phys}$). Once this is determined, we then keep $\epsilon_0$ fixed within $R_{QZM}$, and keep $\epsilon$ as a variable within the moment equation. So long as $\epsilon > \epsilon_0$, we preserve the asymptotic requirements of the OPPQ formalism. Implementing the above process for $\epsilon_0$, we find that it corresponds to the first significant figure for the (eventual) physical energy. The data in Tables 2-5 are generated on this basis.

To generate the orthonormal polynomials we must first define an ordered sequence to the  nonnegative coordinate pairs $\{(m_i,n_i)| 0 \leq m_i,n_i\}$. This sequence must be efficiently chosen relative to the missing moment structure. The most natural choice is in a progression based on their antidiagonal sum: $(0,0)_0, (1,0)_1, (0,1)_2, (2,0)_3, (1,1)_4, (0,2)_5,\ldots$.The orthonormal polynomials for $R_{QZM}(\xi,\eta)$ are defined by ${\cal P}_I(\xi,\eta) \equiv \sum_{i=0}^I\Xi_i^{(I)} \xi^{m_i} \eta^{n_i}$, and their coefficients satisfy the relations $\sum_{i=0}^I\sum_{j=0}^J \Xi_i^{(I)} {\cal W}_{i,j} \Xi_j^{(J)} = \delta_{I,J}$ where ${\cal W}_{i,j}\ \equiv w_{QZM}(m_i+m_j,n_i+n_j)$. The coefficients are then obtained through a Cholesky decomposition of ${\cal W}$ [29].

Assembling all the OPPQ-BM components we have the following. 
The OPPQ-BM expansion takes on the form
\begin{eqnarray}
\Phi(\xi,\eta) = \sum_{I=0}^\infty c_I \ P_I(\xi,\eta) \ R_{QZM}(\xi,\eta),
\end{eqnarray}
and the projection coefficients become (i.e. $c_I = \langle P_I|\Phi\rangle$) 
\begin{equation}
c_I = \sum_{i=0}^I \Xi_i^{(I)} u(m_i,n_i),
\end{equation}
or
\begin{equation}
c_I = \sum_{\ell=0}^{m_s(I)}\Lambda^{(I)}_{\epsilon,\ell}u_\ell,
\end{equation}
where $m_s(I)$ is the missing moment order required to generate $c_I$, and 
\begin{equation}
\Lambda^{(I)}_{\epsilon,\ell} = \sum_{i=0}^I\Xi_i^{(I)}M_E(m_i,n_i,\ell).
\end{equation}

An alternative way to use the above relations is to say that  the first $1+m_s$ missing moments, $\{u_\ell| 0 \leq \ell \leq m_s\}$, can be used to generate all the moments $\{u(m,n)| 0\leq m+n \leq 2m_s+1\}$, through the moment-missing moment relation in Eqs.(50-51). However, these are the moments required in order to generate all the sequentially ordered $c_I$ coefficients satisfying $\{c_I | 0 \leq I \leq  I_{m_s} \equiv (m_s+1)(2m_s+3)-1\}$ in Eqs. (58-59). These $c_I$ coefficients depend on the coefficients of the orthonormal polynomials for the same range of $I$-index values. However, these coefficients require a Cholesky analysis relative to the $R_{QZM}$-moment matrix ${\cal W}_{i,j} =w_{QZM}(m_i+m_j,n_i+n_j)$ where $m_i+m_j+n_i+n_j \leq 2(2m_s+1)$. That is, the generation of the $\Gamma$'s requires $M+N \leq 2(2m_s+1)$. 

The corresponding partial sums, $S_I$, become :
\begin{eqnarray}
S_I(\epsilon,u_\ell) = \sum_{i=0}^I\Big( c_i(\epsilon , u_\ell )\Big)^2, \cr
\hspace{45pt}= \sum_{\ell_1 = 0}^{m_s(I)}\sum_{ \ell_2= 0}^{m_s(I)} u_{\ell_1}{\cal P}_{\epsilon;\ell_1,\ell_2}^{(I)}   u_{\ell_2} ,\cr
\hspace{5pt}{\cal P}_{\epsilon;\ell_1,\ell_2}^{(I)} \equiv \sum_{i=0}^I \Lambda_{\epsilon,\ell_1}^{(i)} \Lambda_{\epsilon,\ell_2}^{(i)},\\
\hspace{5pt} {\cal P}_{\epsilon}^{(I)} = \sum_{i=0}^I {\overrightarrow {\Lambda_\epsilon^{(i)}}}{\overrightarrow {\Lambda_\epsilon^{(i)}}},
\end{eqnarray}
a   symmetric positive definite matrix made up of indiviual semidefinite dyadic matrices. This is because `$I$' is much larger than the dimension ($1+m_s$) of the $\Lambda$-vectors; thereby guaranteeing a sufficient number of independent $\Lambda$ vectors making ${\cal P}^{(I)}_\epsilon$ positive definite.
It is relative to this positive definite matrix that the constrained quadratic form minimization formalism in Sec. 3 is implemented. 

More specifically, the adopted uniform normalization becomes simply $u_0 \equiv 1$, which is expected to be valid (i.e. and not interfere with any symmetry conditions) for the even parity states. That is, the standard normalization $\sum_{\ell =0}^{m_s} u^2_{\ell} =1$ cannot be applied consistently across all ${\cal U}_I$ subspaces introduced in Sec. 3. The uniform normalization, $u_0 = 1$, leads to a constrained quadratic form $S_I(\epsilon,u_0 =1,u_1,\ldots,u_I)$ whose global minimum value over the unconstrained missing moment variables defines the energy dependent function, ${\cal L}_I(\epsilon)$, as defined in Sec. 3 and used in the following section.

\subsection {QZM Numerical Results}

 Table 2 summarizes the  OPPQ-BM results for QZM, including the energy estimates (column three) and the energy bounds (columns four and five),  based on a constrained minimization analysis of the quadratic form ${\cal L}_{I_{m_s}}(\epsilon) \equiv Min_{\overrightarrow u} S_{I_{m_s}}(\epsilon;u_0 =1,u_1,\ldots,u_{m_s})$, as discussed previously in Sec. 3.  The sixth column is the $\epsilon_0$ parameter value used for the reference function weight, as explained earlier. Within the OPPQ-BM formalism, the local minima, $\partial_{\epsilon} {\cal L}_{I_{m_s}}(\epsilon_{I_{m_s}}^{(min)}) = 0$, are the energy estimates given in Table 2, for the expansion orders given in the second column.  The energy bounds given in Table 2 are the result of analysis summarized in Tables 3-5, usually determined at a lower expansion order than that quoted in the second column of Table 2.  We emphasize that the bounds quoted in Tables (2-5) are true bounds for the physical energies. 

For comparative purposes, in Table 2 we quote the energy estimates reported by  Kravchenko et al [28]. The latter results are the higher accuracy estimates in the literature,  yielding twelve-thirteen significant figures for the ground state binding energy, $\epsilon_{gr}$, for magnetic field values $B \leq 4000$. Their results for the first excited state, $\epsilon_1$, vary from twelve significant figures to six, for magnetic field strengths $B \leq 1000$, with no energies reported for higher magnetic fields.  The OPPQ-BM estimates in Table 2 surpass or match their $\epsilon_{gr}$ analysis provided $B \leq 200$.  For $B = 2000$, the OPPQ-BM results for $\epsilon_{gr}$  generate approximately nine of the thirteen significant figures. The only limitation of OPPQ-BM is the computational speed of our computing platform (i.e. MacBook Pro 2.2 GHz/1333MHz). 

For the first excited state, $\epsilon_1$, OPPQ-BM matches or surpasses the accuracy of Kravchenko et al's results  for $B \leq O(200)$.  For $B = 2000$, the OPPQ-BM results for $\epsilon_1$ are compared to those of Schimerczek and Wunner [27]; while for $B= 10^4$ we also compare both states to their B-spline analysis results. The ground state results manifest faster convergence than the first excited state. The generated OPPQ-BM bounds are modest, at these higher magnetic field strengths, given the higher expansion orders required for implementing OPPQ-BM. Tighter bounds would be generated on a faster computer platform, or through an alternate choice to the MER representation chosen here. These possibilities are currently under investigation.  

The ${\cal B}$ values cited in Tables 3-5 are the corresponding values for the quadratic form at the extremum energy values: ${\cal B}_{I_{m_s}} ={\cal L}_{I_{m_s}}(\epsilon_{I_{m_s}}^{(min)})$.  These define a sequence that is positive, increasing, and bounded from above (i.e. $0 < {\cal B}_{I_{m_s}} <  {\cal B}_{I_{m_s+1}} < \ldots < {\cal L}_{\infty}({\epsilon_{phys}}) < \infty$). If this sequence is sufficiently fast converging (as is the case for most of the energies in Tables 3-5), then one can, empirically,  determine a coarse upper bound, ${\cal B}_U$, from the generated sequence elements up to some, relatively low, expansion order:  $\{{\cal B}_{I_{m_s}}| I_{m_s} < I_{max}\}$.  These ${\cal B}_U$ estimates are given in the last column in Tables 3 - 5; and used at a given order (as cited in Tables 3 - 5) to determine the energy interval satisfying ${\cal L}_{I_{m_s}}(\epsilon) < {\cal B}_U$. The endpoints of this interval are the generated bounds, also quoted in Tables 2-5. 

As an example, in Table 3, for the $B = 2$ magnetic field case summarized in Table 2, the fourth column gives the 
$\{{\cal B}_{I_{m_s}}\}$ sequence for $10 \leq m_s \leq 20$ for the ground state. It is clear that the fifteen decimal place number, ${\cal B}_U = 1.192243533462017$ (i.e. entry in fifth column), is a good upper bound estimate of the true limiting form for this sequence, which is already displaying a convergent behavior at the sixteenth decimal place (i.e. $...{\underline 6}490$). On this basis, we determined the energy interval satisfying this ${\cal B}_U$-upper bound, resulting in the $\epsilon$-bounds quoted in Table 3 and summarized in Table 2: $1.02221390766512894 < \epsilon_{gr} < 1.02221390766512930$.

In some cases, such as that for the first excited state in the magnetic field $ B = 0.2$ in Table 4, we update the ${\cal B}_U$ expression three times, based on improved confidence in the convergent behavior of the ${\cal B}_U$ sequence. This is done at $m_s = 22,24,28$, each yielding the tighter bounds cited, yielding the final bounding result $0.14898667819813574694 < \epsilon_1 < 0.14898667819813574698$. The latter is reported in Table 2 as well.

For large magnetic fields, the expansion order required  to obtain results comparable to those in the literature increases. Results corresponding to $I_{m_s} > O(40)$ requires considerable time (i.e. several hours), with available computing resources (i.e. MacBook Pro, 2.2GHz, 1333 MHz, Intel Core I7 processor).  In order to generate bounds, we need very accurate estimates for  the sequence elements ${\cal B}_{I_{m_s}} ={\cal L}_{I_{m_s}}(\epsilon_{I_{m_s}}^{(min)})$ and in particular, the energy extrema $\partial_{\epsilon} {\cal L}_{I_{m_s}}(\epsilon_{I_{m_s}}^{(min)}) = 0$. The function ${\cal L}_{I_{m_s}}(\epsilon)$ has very large gradients, and computing these numerically is difficult. Fortunately, OPPQ-BM can generate closed form expressions for the derivative, $\partial_{\epsilon}{\cal L}_{I_{m_s}}(\epsilon)$, combined with bisection methods for accurately determining the local extrema in the energy variable, $\partial_{\epsilon}{\cal L}_{I_{m_s}}(\epsilon) = 0$, as given in the Tables [29].  That is, for a given $I_{m_s}$ we determined a coarse energy interval $[\epsilon_1,\epsilon_2]$ such that  $\partial_\epsilon {\cal L}_{I_{m_s}}(\epsilon_1) < 0 < \partial_\epsilon {\cal L}_{I_{m_s}}(\epsilon_2)$. The derivatives are determined in closed form. We then used a bisection method to determine the signature at the midpoint energy value, $\partial_\epsilon{\cal L}_{I_{m_s}}(\epsilon_m)$, where $\epsilon_m = {{\epsilon_1+\epsilon_2}\over 2}$. This determines which of the two $\epsilon_{1,2}$ is updated. The process is repeated until the desired accuracy is achieved,  $O(\epsilon_2-\epsilon_1) < 10^{-N}$. Depending on the size of the magnetic field, the $I_{m_s}$ expansion order, and the computing time involved, we could generate $7 \leq N \leq 20$.  We did this primarily to obtain accurate ${\cal B}_{I_{m_s}}$ values with which to generate accurate and tight eigenenergy bounds. All numbers in Tables 2-5 are accurate to the number of digits given (betraying the computational accuracy of our algebraic method), except in a few cases (i.e. the inaccurate digits are underlined) where  we  terminated our bisection algorithm due to time considerations and because the generated results were adequate.

\begin{table}
\caption{ OPPQ-BM  Estimates  and Bounds for QZM: $ \{+,l_z = 0\}$  }
\centerline{
\begin{tabular}{rllllc}
\hline
$B$ & $m_s$ & $\partial_\epsilon{\cal L}_{I_{m_s}}(\epsilon_{I_{m_s}}^{(min)}) = 0$  &    {\it Lower Bound }  & {\it Upper Bound }  & $\epsilon_0$    \\
\hline
\hline
0.02 & 22 & ${0.509900044089401317}_{gr}$ &   0.509900044089401316 &   0.509900044089401318 &0.5 \\
        &      &$0.509900044089\ [28]$ &                                                        &                                         &                                     \\
\hline
        &   22   &  $0.13362417753479289364_1$      &  $0.13362417753479289$  &0.13362417753479291    &0.1     \\
       &      &  $0.133624177534\ [28]$      &               &    \\
\hline
0.20 & 20 & $0.59038156503476258477_{gr}$ & $0.59038156503476258474$  & $0.59038156503476258480$ &0.5\\
      &        &    $0.590381565035\ [28]$                                           &                                    &                              &                                             \\
\hline
      &  28    &    $0.14898667819813574696_{1}$&      0.14898667819813574694& 0.14898667819813574698  & 0.1 \\
        &        & $0.148986678198\ [28]$           &                           &                         &       \\
\hline
 2 & 20 &  $1.02221390766512912_{gr}$ &1.02221390766512894   &1.02221390766512930 &1.0 \\
    & 		&$1.022213907665\ [28]$ 			&			 &	 							 				&		\\
\hline
 & 34&     $0.1739447059728_1$ &0.1739447059 & 0.1739447069 &0.1  \\
 &		&		0.173944705973 [28]   &                           &                        &                                                      \\
\hline
20 & 24 &  $2.21539851543322_{gr}$ &  $2.2153985154326$ &  $2.2153985154375$ &2.0 \\
  &  &  2.215398515433 [28]&   &   &\\
\hline
  & 44&     $0.22384212729_1$ & 0.223842118 &  0.223842138 &0.2  \\
  &    &     0.223842127 [28]    &  &            &  \\
\hline
200& 44 &  $4.72714511068704_{gr}$ & 4.727145110662  &  4.727145110700 &4.0 \\
   &    &      $4.727145110687 [28]$ &  & \\
\hline
&50&        $0.2689772_1$                                & 0.26895            &   .26920     &0.2     \\
 & &            0.2689682 [28] & &       & \\
\hline
2000 & 46 &  $9.304765094_{gr}$ & 9.30475796875&  9.30476699219 &9.0  \\
         &   & $9.304765082770\ [28]$  &     &   \\
\hline
  &    40       &  $0.313_1$                        &             & & 0.3\\
 & 48 & $0.3091_1$ &                                            & &      \\
   &    &0.30624125 [27] &         &        \\
 \hline
10000 & 40 & $14.140995_{gr} $ & 14.137    & 14.143 &14.0\\
&44&14.1409812 &   &   &     \\
&50 &14.1409730 &  &   &  \\
           &         &     $14.14096855\ [27]$                     &               &                                                                \\
\hline
 & 32 &   $0.39533_1 $          &         & &0.3\\
& 40 & 0.37289 &   & &\\
   &    &0.3277107 [27] &         &       & \\
\hline
\end{tabular}}
\end{table}

\newpage

\begin{table}
\caption{
OPPQ-BM Bound Generation
}
\centerline{
\begin{tabular}{ccccc}
\hline
$B$ & $m_s$ &  $\partial_\epsilon{\cal L}_{I_{m_s}}(\epsilon_{I_{m_s}}^{(min)}) = 0$ &   ${\cal L}_{I_{m_s}}(\epsilon_{I_{m_s}}^{(min)}) = {\cal B}_{I_{m_s}}$  &   ${\cal B}_U >Lim_{I_{m_s}\rightarrow \infty} {\cal B}_{I_{m_s}}$    \\
\hline
\hline
2 & 10 & 1.02221390772094855 &   1.1922435334360495594&      \\
2 & 12 & 1.02221390766605681 &   1.1922435334615863759&      \\
2 & 14 & 1.02221390766515350 &   1.1922435334620060902&      \\
2 & 16 & 1.02221390766512978 &   1.1922435334620161910&      \\
2 & 18 & 1.02221390766512914 &   1.1922435334620164834&      \\
2 & 20 & 1.02221390766512913 &   1.1922435334620164906&   1.192243533462017   \\
2 & 22 & 1.02221390766512912 &   1.1922435334620164908&     \\
\hline
Bounds & 20 & $1.02221390766512894 $  & $   1.02221390766512930 $ &${\epsilon^{(L)} }< \epsilon_{gr} < {\epsilon^{(U)} }$ \\
\hline
\hline
2 & 10 & 0.17399695270803026 &  4.5429526411179302253&  \\
2 & 12 & 0.17395614486028235 &  4.5441787609422369991& \\
2 & 14 & 0.17394659357860413 &  4.5444712797500885268 & \\
2 & 16 & 0.17394494775227519 &  4.5445232793907141839 & \\
2 & 18 & 0.17394473400495420 &  4.5445302258621529769& \\
2 & 20 & 0.17394471069126498 &  4.5445309934906803788& \\
2 & 22 & 0.17394470734748571&   4.5445310986100449725&\\
2 & 24 & 0.17394470639567805&   4.5445311272383815527& \\
2& 26 &  0.17394470608302946 &  4.5445311367219481710& \\
2& 28 &  0.17394470599652031 &  4.5445311393995207444& \\
2& 30 &  0.17394470597699553&   4.5445311400165006985& 4.5445312 \\
2& 32 &  0.17394470597336749&   4.5445311401334793794&  \\
2& 34 &  0.173944705972789{\underline{49}}&   4.544531140152356{\underline{7032}}&  \\
\hline
Bounds & 30 & $0.1739447059$  & $  0.1739447069 $ & ${\epsilon^{(L)} }< \epsilon_{1} < {\epsilon^{(U)} }$ \\
\hline
\hline
\end{tabular}}
\end{table}
\newpage
\begin{table}
\caption{
OPPQ-BM Bound Generation (Continued)
}
\centerline{
\begin{tabular}{cclll}
\hline
$B$ & $m_s$ &  $\partial_\epsilon{\cal L}_{I_{m_s}}(\epsilon_{I_{m_s}}^{(min)}) = 0$ &   ${\cal L}_{I_{m_s}}(\epsilon_{I_{m_s}}^{(min)}) = {\cal B}_{I_{m_s}}$  &   ${\cal B}_U >Lim_{I_{m_s}\rightarrow \infty} {\cal B}_{I_{m_s}}$    \\
\hline
\hline
0.02 & 20 & ${0.509900044089401317}$ &  0.2598901748007018&  \\
        & 22 & ${0.509900044089401317}$  & 0.2598901748007018& 0.25989017480071\\
\hline
Bounds & 22 & 0.509900044089401316 &  0.509900044089401318&${\epsilon^{(L)} }< \epsilon_{gr} < {\epsilon^{(U)} }$ \\
\hline
\hline
0.02 & 20 & 0.13362417753479289817 &  0.17403263429090443146&  \\
        & 22 & 0.13362417753479289364 &0.17403263429090443888 & 0.174032634290905\\
\hline
Bounds & 22 & 0.13362417753479289 &0.13362417753479291 &${\epsilon^{(L)} }< \epsilon_{1} < {\epsilon^{(U)} }$ \\
\hline
\hline
0.2 & 18 &  0.59038156503476258478&  0.33642607127667690598&  \\
        & 20 &0.59038156503476258477 & 0.33642607127667690598 &0.33642607127667692 \\
\hline
Bounds & 20 & 0.59038156503476258474  &0.59038156503476258480 &${\epsilon^{(L)} }< \epsilon_{gr} < {\epsilon^{(U)} }$ \\
\hline
\hline
0.2 & 18 &   0.14898667819813825399& 0.69627363002726127600 &  \\
        & 20 & 0.14898667819813583845 &  0.69627363002727750928& \\
        & 22 & 0.14898667819813575011 &  0.69627363002727810688& 0.6962736300273\\
       & 24 &  0.14898667819813574709 &  0.69627363002727812739& 0.6962736300272785 \\
       & 26 &  0.14898667819813574697&   0.69627363002727812821&  \\
  &      28 &  0.14898667819813574696&   0.69627363002727812824&  0.69627363002727813\\
\hline
Bounds & 22& 0.14898667819813515625& 0.14898667819813593750 &${\epsilon^{(L)} }< \epsilon_{1} < {\epsilon^{(U)} }$ \\
            & 24& 0.14898667819813574&   0.14898667819813575 &\\
            & 28& 0.14898667819813574694& 0.14898667819813574698&\\

\hline
\hline
20 & 18 & 2.21539851545624653778 &  5.17117123763253297433&  \\
    &  22&  2.21539851543364863709 &  5.17117123764170817069 &  \\
    & 24 &  2.21539851543326570420 &                         5.17117123764189485910 & 5.171171237646 \\
    & 26 &  2.21539851543323075{\underline{867}} &    5.171171237641913583{\underline {19}}& \\
    & 28 &  2.21539851543322723275 &      5.17117123764191534218& \\
\hline
Bounds & 24 & 2.2153985154326  &2.2153985154375 &${\epsilon^{(L)} }< \epsilon_{gr} < {\epsilon^{(U)} }$ \\
\hline
\hline
20 & 18 &  0.22387845348680& 54.380271821544418182 &  \\
        & 22 &  0.2238479698250145& 54.391660422765839227 & \\
  &   26 &      0.2238429228787462 & 54.393555147477881277 & \\
& 36&     0.223842132100{\underline {4028}}& 54.39385358621623{\underline {4652}} & \\
 &40 &    0.2238421283199{\underline {730}} & 54.3938550076989{\underline {39964} }& 54.39386\\
&44&         0.2238421272{\underline{868750}}&      54.393855391{\underline{128120877 }}& \\ 
\hline
Bounds & 40 & 0.223842118 &  0.223842138 & ${\epsilon^{(L)} }< \epsilon_{1} < {\epsilon^{(U)} }$ \\
\hline
\hline
\end{tabular}}
\end{table}

\begin{table}
\caption{\label{tab1}
OPPQ-BM Bound Generation (Continued)
}
\centerline{
\begin{tabular}{cclll}
\hline
$B$ & $m_s$ &  $\partial_\epsilon{\cal L}_{I_{m_s}}(\epsilon_{I_{m_s}}^{(min)}) = 0$ &   ${\cal L}_{I_{m_s}}(\epsilon_{I_{m_s}}^{(min)}) = {\cal B}_{I_{m_s}}$  &   ${\cal B}_U >Lim_{I_{m_s}\rightarrow \infty} {\cal B}_{I_{m_s}}$    \\
\hline
\hline
200 & 20 & 4.72714521527550505 & 29.642742337435354224 &  \\
        & 26 &4.72714511208413433 & 29.642742415692537444& \\
        & 32 &4.72714511077668{\underline {309}} & 29.642742416707460743 & \\
        & 40 &4.72714511068894{\underline {043}}&               29.642742416764079{\underline {093}} &29.6427424168 \\
           & 44 &4.7271451106870{\underline{4219}}&          29.642742416765467{\underline{663}}& \\
\hline
Bounds & 40 & 4.727145110662  &  4.727145110700&${\epsilon^{(L)} }< \epsilon_{gr} < {\epsilon^{(U)} }$ \\
\hline
\hline
200 & 20 & 0.2717245007705{\underline{4637}} &  315.72878015439875284&  \\
        & 26 & 0.269586502689{\underline {49568}} &318.78457842049146957 \\
 & 32 &    0.269138248515097{\underline{80}} & 319.42980234551242656 \\
&36 &      0.2690492882396{\underline{5602}} & 319.55775248188720730\\
&40  &     0.2690098{\underline{7548828125}}&  319.6143280{\underline {0806839988}}\\
&44 &      0.2689906{\underline {7749023438}} &        319.6418452{\underline{9691915874}}&319.72\\
&50&          0.268977{\underline{21375}} &               319.6607{\underline{7073116848068}}\\
\hline
\hline
Bounds & 44 &  0.26895 & .26920   &${\epsilon^{(L)} }< \epsilon_{1} < {\epsilon^{(U)} }$ \\
\hline
2000 & 10 & 9.3105417878936{\underline{85978}} & 208.71773742661461146 &  \\
        & 20 & 9.3048296437300{\underline{30512}}&                                                          208.73337152719312079  & \\
   & 30 &9.3047680878{\underline{49020958}}   &208.73352972294715685 &  208.74  \\
 &40 &   9.304765187{\underline{114477157}} & 208.7335371512528{\underline{4458}}& 208.734\\
&46& 9.30476509{\underline{3765625 }}&208.733537394{\underline{28614152}}& \\
\hline
Bounds & 30& 9.3047  & 9.3049&${\epsilon^{(L)} }< \epsilon_{gr} < {\epsilon^{(U)} }$ \\
 &40&9.30475796875&  9.30476699219  \\
\hline
2000 &20 &         0.360665443859{\underline {215884}} &  2952.8607929709714042  &  \\
 &30 &         0.325516013{\underline{830900192}}&    3399.1752494826152{\underline{309}}&  \\
 &40 &  0.313                        &       3593.48      & \\
&48 &     0.3091{\underline{4453125}}  & 3655.84{\underline{69805732769106}}&\\
\hline
\hline
\hline
10000 & 10 & 14.20793218107{\underline{684879}}  &  840.64580049062185769 &  \\
          & 20 & 14.143328544542{\underline{23707}} &                         841.01755510169068617 & \\
          & 30 & 14.141141024725{\underline{06056}} &                         841.03035832298493362 & 841.07 \\
    & 44  &   14.1409812{\underline {6953125000}}  &    841.03127064{\underline{101628413}}& \\
&50 &     14.1409730{\underline{3432617188}} &841.0313166336{\underline{2539300}}&\\
\hline
\hline
Bounds & 30 & 14.137  & 14.143&${\epsilon^{(L)} }< \epsilon_{gr} < {\epsilon^{(U)} }$ \\
\hline
\hline
10000&20& 0.464734628072{\underline{710620}} & 10372.371983390232675 & \\
         &24& 0.4328929819{\underline{89880354}} &                                    11068.764546826187807 & \\
        &32&  0.3953300335{\underline{24078317}}&                                     12200.468986908568427& \\
        &40&  0.3728917{\underline{50335693359}}&                                     13077.992260736{\underline{788789}}& \\
\hline
\end{tabular}}
\end{table}

\section{Conclusion}
We have presented a new algebraic eigenenergy bounding method applicable to any multidimensional quantum system admitting a moment equation representation. 
The method applies to either bosonic or fermionic systems. The importance of moment equation representations is underscored through this formalism. The availability of high accuracy algebraic software, such as Mathematica, recommends the OPPQ-BM formalism for generating converging bounds to sensitive SCSPS type systems. So long as the coarse upper bound, ${\cal B}_U$, to the limit of the converging positive sequence ${\cal L}_i(E_i^{(min)})$, can be reliably established, the OPPQ-BM bounds can be generated. If this is not the case, the OPPQ-BM eigenenergy approximants should be of sufficient high accuracy for tackling any problem. If neither is the case, then the EMM bounding formulation (anticipating its near term extension to multidimensional excited states) should provide an alternative bounding strategy. 

We also note that there are many different MER representations, and weight-reference functions, for a given problem. We are considering these alternate strategies for improving the results presented here, particularly for the excited states.

Finally, in principle, we can reverse the roles of the configuration space and Fourier space. Let $\hat{\Psi}(k) = {1\over{\sqrt{2\pi}}}\int dx \ exp(-ikx)\Psi(x)$. This will be analytic in $k$ for configuration space physical solutions that are sufficiently exponentially decaying. If the inverse Fourier transform is also sufficiently exponentially decaying then $\Psi(x) = {1\over{\sqrt{2\pi}}}\int dk \ exp(ixk){\hat\Psi}(k)$ will also be analytic in $x$. Its power series expansion involves the $k$-space power moments:  $\Psi(x) = {1\over{\sqrt{2\pi}}} \sum_{j=0}^\infty {{(ix)^j}\over {j!}}\int dk \ k^j {\hat\Psi}(k)= \sum_j c_j(E,c_0,c_1) x^j$. What is needed is an OPPQ analysis within the Fourier space: 
$\hat{\Psi}(k) = \sum_{j=0}^\infty {{\cal P}_j(k)} {\cal R}(k)$. However, this approach is harder to implement because the asymptotic properties of the physical solutions in the Fourier space are not easy to calculate [30]; thereby making an OPPQ analysis more difficult. Only for configuration space potentials where the discrete state asymptotic form is Gaussian, and where the Fourier asymptotic form is also Gaussian, can the generation of ${\cal R}(k)$ yield any readily useful results. 
\section{Acknowledgement}
The author is appreciative of Dr. John R. Klauder and Dr. Daniel Bessis for inspiring comments and insight received over many decades that led to the realization of this work. The technical assistance of Dr. Maribel Handy is greatly appreciated.

\newpage

\section{References}
\noindent [1] Bender C M and Orszag S A 1999 {\it Advanced Mathematical Methods for Scientists and Engineers} (New York: Springer)\\
\noindent [2] Morse P M and Feshbach H 1953 {\it Methods of Theoretical Physics} (New York: MaCraw-Hill Book Co., Inc.)\\
\noindent [3] Temple G  1928 {\em Proc. R. Soc. A Math. Phys. Eng. Sci.} {\bf 119} 276 \\
\noindent [4] Barta J 1937 {\em C. R. Acad. Sci. Paris} {\bf 204} 472 \\
\noindent [5] Handy C R and  Bessis D 1985 {\em Phys. Rev. Lett.} {\bf 55} 931 \\
\noindent [6] Handy C R, Bessis D, Sigismondi G, and Morley T D 1988 {\em Phys. Rev. A} {\bf 37} 4557 \\
\noindent [7] Handy C R, Bessis D, Sigismondi G, and Morley T D 1988 {\em Phys. Rev. Lett.} {\bf 60} 253 \\
\noindent [8] Handy C R, Appiah K and Bessis D 1994 {\em Phys. Rev. A} {\bf 50} 988\\
\noindent [9] Lasserre J-B  {\it { Moments, Positive Polynomials and Their Applications}} (London: Imperial College Press 2009)\\
\noindent [10] Boyd S and Vandenberghe L 2004 {\it Convex Optimization} (New York: Cambridge University Press)\\
\noindent [11] Chvatal V 1983 {\it Linear Programming} (Freeman, New York)\\
\noindent [12] Shohat J A and  Tamarkin J D, 1963 {\it The Problem of Moments} (American Mathematical Society, Providence, RI) \\
\noindent [13] Reed M and Simon B 1978 {\it Methods of Modern Mathematical Physics} (New York: Academic)\\
\noindent [14] Handy C R 1987 {\em Phys. Rev. A} {\bf 36}, 4411 \\
\noindent [15] Handy C R 2001 {\em J. Phys. A} {\bf 34} L271\\
\noindent [16] Handy C R 2001 {\em J. Phys. A} {\bf 34} 5065\\
\noindent [17] Le Guillou J C and Zinn-Justin J 1983  {\em Ann. Phys. (N.Y.)} {\bf 147} 57 \\
\noindent [18] Barnsley M F and Demko S G 1985 {Pro. R. Soc. London, Sec A} {\bf 399} 243\\
\noindent [19] Handy C R and Mantica G 1990 {\it Physica D} {\bf 43} 17\\
\noindent [20] Bessis D and Demko S 1991 {\it Physica D} {\bf 47} 427\\
\noindent [21] Klauder J R 2019 {\it arXiv} 1912.08047\\
\noindent [22] Delabaere E and Trinh D T 2000 {\it J. Phys. A Math. Gen} {\bf 33} 8771\\
\noindent [23] Handy C  R and Lee P 1991 {\em J. Phys. A. Math. Gen.} {\bf 24} 1565\\
\noindent [24] Handy C R and Vrinceanu D 2013 {\em J. Phys. A: Math. Theor. }  {\bf 46} 135202   \\
\noindent [25] Amore P and Fernandez F M 2012 {\em J. Phys. B: At. Mol. Opt. Phys.} {\bf 45} 235004  \\
\noindent [26] Handy C R and Vrinceanu D 2013 {\em J. Phys. B: At. Mol. Opt. Phys.} {\bf 46} 115002 \\ 
\noindent [27] Schimerczek C and Wunner G 2014 {\em Comp. Phys. Comm.} {\bf 185}  614 \\
\noindent [28] Kravchenko Y P, Liberman M A,  and Johansson B  1996 {\em Phys. Rev. A} {\bf 54} 287 \\
\noindent [29] Handy C R 2020 {TSU-Preprint}\\
\noindent [30] Handy C R, Vrinceanu D, Marth C B and Gupta R 2016 {\em J. Phys. A: Math. Theor.} {\bf 49}, 145205\\

\end{document}